% ...
\documentclass[twocolumn,showpacs,preprintnumbers,amsmath,amssymb,showkeys]{revtex4}

\usepackage{graphicx}
\usepackage{dcolumn}
\usepackage{bm}
\usepackage{enumerate}
\usepackage{algorithm}
\usepackage{mathrsfs}
\usepackage[noend]{algorithmic}

\newtheorem{theorem}{Theorem}[section]

\newtheorem{lemma}[theorem]{Lemma}

\newtheorem{conjecture}[theorem]{Conjecture}
\newenvironment{proof}[1][Proof]{\textbf{#1.} }{\ \rule{0.5em}{0.5em}}

\def\satur{\mathop{\rm Satur}\nolimits}
\newcommand{\supp}[1]{\text{supp}\thinspace #1}
\newcommand{\indicator}[1]{1\hspace{-1.0mm}{\text{\rm l}}_{\{#1\}}}
\newcommand{\ent}[1]{{\rm Ent}\{#1\}}

\begin{document}

\preprint{APS/123-QED}

\title{Inference and Optimal Design\\for Nearest-Neighbour Interaction Models}

\author{A.\ Iu.\ Bejan}
\email{Andrei.Bejan@cl.cam.ac.uk}
\altaffiliation[also at ]{Computer Laboratory, University of Cambridge}
 \homepage{http://www.cl.cam.ac.uk/~aib29}
\author{Gavin J.\ Gibson}
\author{Stan Zachary}
\affiliation{%
Department of Actuarial Mathematics and Statistics and the Maxwell Institute for Mathematical Sciences,\\
Heriot-Watt University, Edinburgh, EH14 4AS, UK}

\date{\today}

\begin{abstract}
We consider problems of Bayesian inference for a spatial epidemic on
a graph, where the final state of the epidemic corresponds to bond
percolation, and where only the set or number of finally infected
sites is observed.  We develop appropriate Markov chain Monte Carlo
algorithms, demonstrating their effectiveness, and we study problems
of optimal experimental design.  In particular, we demonstrate that
for lattice-based processes an experiment on a sparsified lattice
can yield more information on model parameters than one conducted on
a complete lattice. We also prove some probabilistic results about
the behaviour of estimators associated with large infected clusters.
\end{abstract}

\maketitle

\section{\label{sec:Introduction}Introduction}
Many real-world phenomena can be modelled by random graphs, or more
generally, by dynamically changing random graphs. Specifically,
host-pathogen biological systems that may combine primary and
nearest-neighbour or long-range secondary infection processes can be
efficiently described by spatio-temporal models based on random
graphs evolving in time~\cite{Gibson&et&al&2006}.

Although a continuous observation of an epidemic is not always
possible, a spatial `snapshot' may provide one with some, albeit
highly incomplete, knowledge about the epidemic. In terms of the
model this knowledge results in a random graph realised in some
metric space. Moreover, under some circumstances it is not even
possible to observe some or all of the edges of such a random
graph---all one would know then are the vertices which correspond to
the infected sites, i.e.\ to those sites which interacted as a
result of the evolution of the process under consideration.

One particular application refers to the colonisation of susceptible
sites, such as seeds or plants grown on a lattice, by virus, fungal,
or bacterial pathogens with limited dispersal abilities. A typical
example is the spread of infections through populations of seedlings
by the fungal pathogen, \textit{Rhizoctonia solani} K\"{u}hn. This
economically-important pathogen is wide spread with a remarkably
wide host range \cite{Chase1998}. In addition to its intrinsic
economic importance, it has been extensively used as an experimental
model system to test epidemiological hypotheses in replicated
microcosms
\cite{Gibson&Kleczkowski&Gilligan2004,Otten&Bailey&Gilligan2004} and
to study biological control of pathozone behaviour by an
antagonistic fungus and disease dynamics~\cite{Bailey&Gilligan1997}.
Transmission of infection between plants occurs by mycelial growth
from an infected host, with preferential spread along soil
surfaces---hence the missing information about the structure of
interactions.

The spread of infections with limited dispersal abilities among
plants can be viewed as a spatial \textit{SIR}
(susceptible$\thinspace\rightarrow\thinspace$infective$\thinspace\rightarrow\thinspace$removed)
epidemic with nearest-neighbour secondary infections and removals,
and can be related to a percolation process on a finite static
network~\cite{Gibson&et&al&2006,Bailey&Otten&Gilligan2000}.
Percolation has also been widely used to model various phenomena of
disordered media in physics, chemistry, biology, engineering and
materials science~\cite{Sahimi1994,Frisch&Hammersley}, and also
forest fires~\cite{Zhang,Cox&Durrett}.

Bayesian estimation for percolation models of disease spread in
plant populations in the context of the spread of
\textit{Rhizoctonia solani} has been presented in
\cite{Gibson&et&al&2006}. In \cite{Bailey&Otten&Gilligan2000} the
spread of this soil-borne fungal plant pathogen among discrete sites
of nutrient resource is studied using simple concepts of percolation
theory; a distinction is made between invasive and noninvasive
saprotrophic spread. The authors of
\cite{Gibson&et&al&2006,Bailey&Otten&Gilligan2000} formulated
statistical methods for fitting and testing percolation-based,
spatio-temporal models that are generally applicable to biological
or physical processes that evolve in time in spatially distributed
populations. Estimation of spatial parameters from a single snapshot
of an epidemic evolving on a discretised grid under the assumption
that fundamental spatial statistics are near equilibrium is studied
in~\cite{Keeling&Brooks&Gilligan2004}.

The difficulties in performing inference for these models in the
presence of observational uncertainty or incomplete observations can
be overcome to an extent by employing a Bayesian approach and modern
powerful computational techniques---mainly Markov chain Monte Carlo
(McMC), e.g.~\cite{Gibson}. McMC methods often offer important
advantages over existing methods of analysis. Particularly, they
allow a much greater degree of modelling flexibility, although the
implementation of McMC methods can be problematic because of
convergence and mixing difficulties which arise due to the amount
and nature of missing data.

An aspect which has received little attention in the context of the
described models is that of experimental design. Statisticians have
investigated the question of experimental design in the Bayesian
framework (see \cite{Chaloner_Verdinelli1995} for a review). The
work of M\"{u}ller and others~\cite{Muller1999,Verdinelli1992}
examined the ways of identifying designs that maximise the
expectation of a utility function.

The goal of this paper is to study inference and optimal design
problems for a model which is obtained by taking a final snapshot of
nearest-neighbour disease spread spatio-temporal dynamics and
relating it to the percolation process on a regular grid. We use the
utility-based Bayesian framework for this purpose and discuss
generic issues that arise in this context.

The paper is arranged as follows. In Section 2 we describe the
percolation model considered and describe the partial sampling
scenarios that form the focus of the paper. In Section 3 we describe
how, by embedding the problem in a framework of Bayesian inference,
it is possible to estimate likelihood functions for percolation
models under the respective sampling scenarios using Markov chain
Monte carlo (McMC) methods.  Moreover we prove some theoretical
results regarding the limiting properties of the likelihood for the
case when only cluster size is observed, and elicit interesting
connections with percolation thresholds. In Sections 5 and 6 we turn
our attention to the question of experimental design and, in
particular, the identification of experimental lattices that can
maximise the information on the percolation probability yielded by
an experimental process observed on it, when only the vertex set of
the connected cluster is observed.  We demonstrate that the optimal
lattice may be one which includes regions which have been sparsified
by removal of certain vertices and edges.  These results have
important implications for the design of experiments on, for
example, fungal pathogen spread in populations of agricultural
plants, where hosts are typically arranged on a lattice.

\section{\label{sec:Main}Nearest-neighbour interaction model and percolation}

\subsection{Model description}
Disease spread as a result of (typically) short-range contact
between, for example, plants can be modelled as a transmission
process on an undirected graph.  Nodes, or vertices, of the graph
correspond to possible locations of plants, and edges of the graph
link locations which are considered to be \emph{neighbours}.  In a
classical \textit{SIR} model each node, or vertex, of the graph is
in one of three states: either it is occupied by a \emph{healthy},
but susceptible, plant (state~\textit{S}), or it is occupied by an
\emph{infected} and infectious plant (state~\textit{I}), or finally
it is \emph{empty}, any plant at that location having died and thus
being considered removed (state~\textit{R}).  A plant at node $i$,
once infected (or from time $0$ if initially infected), remains in
the infected state~\textit{I} for some random lifetime $\tau_i$
after which it dies, so that node $i$ then remains in the empty
state~\textit{R} ever thereafter. During its infected lifetime the
plant at node $i$ sends further infections to each of its
neighbouring nodes $j$ as a Poisson process with rate $\lambda_{ij}$
(so that the probability that an infection travels from $i$ to $j$
in any small time interval of length $h$ is $\lambda_{ij}h+o(h)$ as
$h\to0$ while the probability that two or more infections travel in
the same interval is $o(h)$ as $h\to0$); any infection arriving at
node $j$ changes the state of any \emph{healthy} plant there to
\emph{infected}, and otherwise has no effect.  All lifetimes and
infection processes are considered independent.  The initial state
of the system is typically defined by one or more nodes being
occupied by infected plants, the remaining nodes being occupied by
healthy plants. The epidemic may \emph{die out} at some finite time
at which the set of infected nodes first becomes empty, or, on an
infinite graph only, it is possible that it may continue forever.

Thus, for any infected node $i$, the event~$E_{ij}$ that any
neighbouring node $j$ receives at least one infection from node~$i$
has probability $p_{ij}=1-\mathbb{E}[\exp(-\lambda_{ij}\tau_i)]$
(where $\mathbb{E}$ denotes expectation).  Note that, for any given
node $i$, even though the infection processes are independent, the
events~$E_{ij}$ are themselves independent if and only if the random
lifetime $\tau_i$ is a constant.  We now suppose that this is the
case and that furthermore, for all ordered pairs $(i,j)$ of
neighbours, we have $p_{ij}=p$ for some probability $p$. Suppose
further that it is possible to observe neither the time evolution of
the epidemic nor the edges of the graph by which infections travel,
but only the initially infected set of nodes and the set of nodes
which are at some time infected and thus ultimately in the empty
state~\textit{R}.  It is then not difficult to see, and is indeed
well known~\cite{Grassberger1983,Kuulasmaa&Zachary}, that the
epidemic may be probabilistically realised as an unoriented
\emph{bond percolation} process on the graph in which each edge is
independently \emph{open} with probability $p$ (or \textit{closed}
with probability $1-p$), and in which the set of nodes which are
ever infected consists of those nodes reachable along open
\emph{paths} (chains of open edges) from those initially infected.
(Note that the ability to use an \emph{unoriented} bond percolation
process requires both the assumptions that the above events $E_{ij}$
are independent and that $p_{ij}=p_{ji}$ for all $i,j$; in the
absence of \emph{either} of these assumptions one would in general
need to consider an oriented process with the appropriate dependence
structure~\footnote{Non-constant lifetime distributions $\tau_i$ can
  similarly lead to other interesting percolation processes.  For,
  example, site percolation may be approximated arbitrarily closely by
  a lifetime distribution which with some sufficiently small
  probability takes some sufficiently large value, and which otherwise
  takes the value zero.}.)

In the present paper we consider the epidemic to take place on some
subset $P$ of $\mathbb{Z}^d$, where we allow $P=\mathbb{Z}^d$ as a
possibility. Two sites (nodes) are considered \emph{neighbours} if
and only if they are distance $1$ apart and thus $\mathbb{Z}^d$
induces a graph, called the \textit{$d$-dimensional cubic lattice}
$\mathbb{L}^d=(\mathbb{Z}^d,\mathbb{E}^d)$, with the set
$\mathbb{E}^d$ of edges connecting neighbours. Thus in the case
$P=\mathbb{Z}^2$, for example, each node has 4 neighbours. Notice
that the set $P$ also induces a graph with respect to the lattice
$\mathbb{L}^d$---we shall denote this graph $\Pi$. This may be
considered as a model for \emph{nearest-neighbour} interaction.

We assume furthermore that initially there is a single infected
site, and that all other sites in $\Pi$ are occupied by healthy
individuals.

\subsection{Incomplete observations}
The probability $p$ introduced above is considered to be unknown,
possibly depending, through the experimental design, on other
parameters (for example, on the spacing of the grid). We consider
inference under each of the following two scenarios:
\renewcommand{\labelenumi}{(\roman{enumi})}
\begin{enumerate}
\item the observations consist of the set of sites which are ever
infected, so that the routes by which infections travel are not
observed; note that, in terms of the bond percolation process, this
corresponds to one's knowledge of the connected component containing
the initially infected site---the location of this site within the
component not being relevant to inference for $p$ (see below);
\item all that is
observed is the \emph{size} of the set of sites which are ever
infected.
\end{enumerate}
We refer further to the former of these two scenarios as
$\mathcal{S}1$ and to the latter scenario as $\mathcal{S}2$.


Since for any given set of initially infected sites, the
distribution of the set of ever-infected sites in an \textit{SIR}
epidemic with constant infectious times is the same as for the
corresponding unoriented bond percolation process (with the same
initial set of the nodes involved), a final snapshot of such
epidemic can be seen as an open cluster of the corresponding
percolation process.

\begin{figure}[ptb]
\centering
\includegraphics[width=1.8in]{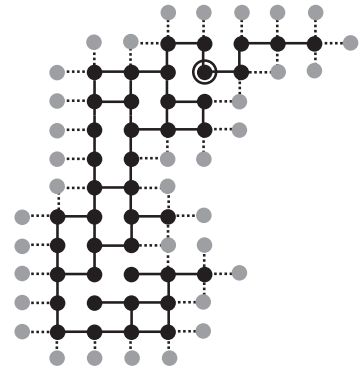}
\caption{An open cluster (black solid dots) containing the origin (a
black dot in a circle) as a result of percolation simulation on
$\mathbb{L}^2$. Here the bond percolation probability $p$ was taken
to be $0.478$; the solid segments represent open bonds. The open
cluster can be seen as a finite outbreak of an epidemic with
constant infectious periods and neighbour-to-neighbour infection
intensity spread rate $\lambda_{ij}\approx 0.65$ (since
$0.478=1-e^{-0.65}$). The dotted lines depict directions along which
infection did not spread (from black to grey dots); thus, grey dots
depict individuals which remain healthy and the dotted lines
represent those bonds that must be closed given the knowledge of the
cluster set.}
\label{fig:outbreak}%
\end{figure}

Figure~\ref{fig:outbreak} depicts an open cluster obtained by
simulation of percolation process on the integer lattice
$\mathbb{Z}^2$ for $p=0.478$. This connected component containing
the origin can be seen as a final (and finite) outbreak of an
\textit{SIR} epidemic process of the kind discussed above. The
origin (or, indeed, any other vertex of the open cluster) may be
considered to be the site where the initially inoculated individual
has been placed. Clearly, the realised bond structure is not the
only possible way which results in the site configuration from
Figure~\ref{fig:outbreak}. However, the distribution of this site
configuration as an extinct \textit{SIR} epidemic coincides with
that of the corresponding unoriented bond percolation process.

\section{Parameter estimation}

\subsection{Scenario $\mathcal{S}1$}
Let $\Pi$ be a (proper or improper) subgraph of $\mathbb{L}^d$ which
contains the origin and let $\mathcal{C}$ be an open cluster of a
percolation process on $\Pi$ containing the origin. The set of nodes
$\mathcal{C}$ represents a snapshot of an extinct outbreak of our
spatial \textit{SIR} epidemic evolved on $\Pi\subseteq\mathbb{L}^2$.

Let us introduce some additional notions. Let $G=(V,E)$ be a locally
finite graph and let $G'=(V',E')$ be a subgraph of $G$ (that is
$V'\subseteq V$ and $E'\subseteq E$). By the \textit{saturation} of
the graph $G'$ with respect to $G$ we understand the graph
$\tilde{G}=(\tilde{V},\tilde{E})$ such that
$$
\tilde{V}=V'\text{ and }\tilde{E}=\{(x,y)\,|\,x,y\in
V'\,\&\,(x,y)\in E\}.
$$
We denote the saturation of $G'$ with respect to $G$ by $\satur_GG'$
or, in cases when it is clear from the context with respect to what
graph the saturation takes place, by $\satur G'$. A graph $G'$ whose
saturation coincides with itself is called a \textit{fully saturated
graph}. For example, the fully saturated graph (with respect to
$\mathbb{L}^2$) is obtained from the graph depicted in
Figure~\ref{fig:outbreak} by connecting all pairs of neighbouring
black sites (according to the 4-neighbourhood relationship). Note
that the operation of saturation may also be applied solely to a
subset of vertices of the original graph, since it does not make use
of the edges of the subgraph-operand.

To distinguish between the boundary points of a graph and their
neighbours, which are not in the graph, we introduce the notions of
the \textit{surface} and the \textit{frontier} of the graph (again,
with respect to another graph). Let us denote by $\partial G$ the
surface of $G$ in $\Pi$, $G\subseteq\Pi$, that is to say the set
$$\partial G:=\{x\in G: \, \exists y\in\Pi\backslash G \text{ s.t.\ }x\text{ and }y\text{ are neighbours in }
\Pi\},$$ and by $\Gamma_G$ the frontier of $G$ in $\Pi$, i.e.\ the
set $\partial (\Pi\backslash G)$.

In order to identify the likelihood function we introduce the set
$\mathcal{G}(\mathcal{C})$ of all \textit{connected} subgraphs of
$\Pi$ with $\mathcal{C}$ as a vertex set. Note that the set
$\mathcal{G}(\mathcal{C})$ is necessarily nonempty.  For each
$G\in\mathcal{G}(\mathcal{C})$ the number of edges between the
vertices of the graph $G$ and the elements of its frontier
$\Gamma_{G}$ is constant---we denote it by $w_\mathcal{C}$. Finally,
we denote the total number of edges in $G$ by $e(G)$.

The probability that $\mathcal{C}$ represents the set of
ever-infected sites and that the edges of $G$ correspond to those
routes along which the infection travelled is
$$
\mathbb{P}_p(G)=p^{e(G)}(1-p)^{e(\satur{\mathcal{C}})-e(G)+w_\mathcal{C}},
$$
and the likelihood function associated with the observed set
$\mathcal{C}$ of ever-infected sites is given by

$$
\mathcal{L}(p)=\mathbb{P}_p(\mathcal{C})=\sum\limits_{G\in\mathcal{G}(\mathcal{C})}\mathbb{P}_p(G).
$$

Hence, under assumption of a uniform prior for $p$, its posterior
distribution $\pi(p\,|\,\mathcal{C})$ is a mixture of beta
distributions:
$$
\pi(p\,|\,\mathcal{C})\propto
\sum\limits_{k}r(k)\thinspace\text{Beta}\left(k+1,e(\satur{\mathcal{C}})-k+w_\mathcal{C}+1\right),
$$
where
$$r(k):=\#\{G\in\mathcal{G}(\mathcal{C})\,|\, e(G)=k\}.$$

It is not feasible to calculate $\pi(p\,|\,\mathcal{C})$ in the
above form, since it is hard to enumerate all corresponding graphs
and, thus, to calculate efficiently the coefficients $r(k)$. We
describe therefore an McMC algorithm that allows one to sample from
the distribution $\pi(p\,|\,\mathcal{C})$ under the uniform prior on
$p$, that is, effectively, to estimate the likelihood function of
$p$.

Our Markov chain explores the joint space
$(0,1)\times\mathcal{G}(\mathcal{C})$ of possible values for $p$ and
graphs from $\mathcal{G}(\mathcal{C})$. The stationary distribution
of the chain is the joint posterior distribution of $p$ and
$G\in\mathcal{G}(\mathcal{C})$. The description of the chain is
given in Algorithm~\ref{alg:MCMC_S1} (Appendix~\ref{appendix:mcmc}).
This Markov chain explores the set of all connected graphs
$\mathcal{G}(\mathcal{C})$ by simply deleting or adding an edge from
the current graph preserving the connectivity of the given site
configuration~$\mathcal{C}$.

\begin{figure}[ptb]
\centering
\includegraphics[width=3.2in]{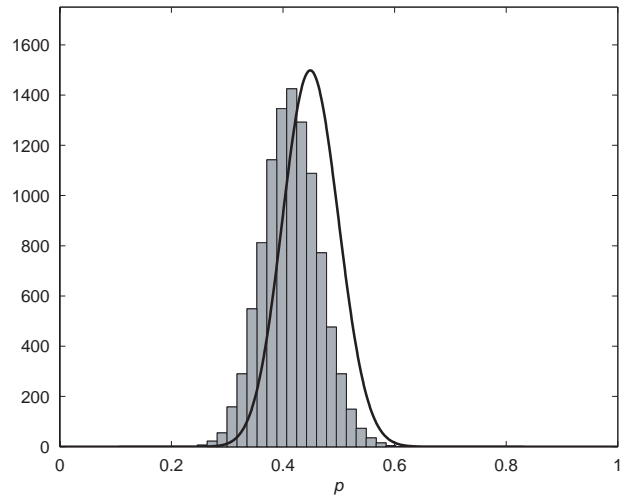}
\caption{The solid line corresponds to the likelihood function
evaluated for the complete information (both the sites and edges are
known) on the cluster $\mathcal{C}$ from Figure~\ref{fig:outbreak}.
The histogram is based on a large sample drawn from the McMC applied
to the site configuration $\mathcal{C}$ (nodes only).}
\label{fig:post_n_likelihood}%
\end{figure}

We apply Algorithm~\ref{alg:MCMC_S1} to the site configuration
$\mathcal{C}$ from Figure~\ref{fig:outbreak} (black dots only).
Figure~\ref{fig:post_n_likelihood} depicts the likelihood function
of the model parameter for the complete observations (i.e.\ nodes
and edges of the cluster) and a histogram of the sample obtained
from the posterior distribution $\pi(p\,|\,\mathcal{C})$ when the
prior distribution is uniform on the interval $(0,1)$.

Mixing properties of Markov chains described by
Algorithm~\ref{alg:MCMC_S1} are briefly discussed in
Appendix~\ref{appendix:Alg12_and_S12}.

\subsection{Scenario $\mathcal{S}2$}
\label{sec:S2}

Under this scenario only the size $n$ of the outbreak of our
\textit{SIR} epidemic evolving on $\Pi=\mathbb{L}^d$ is given.

Let $\mathcal{G}_n$ be the set of all possible connected graphs on
$n$ vertices including the origin. These graphs represent the
outbreaks of the size $n$ and we distinguish all isomorphic graphs
which have different locations or orientations.

We denote the number of edges of $\Pi$ between the vertices of the
graph $G\in\mathcal{G}_n$ and the vertices of its frontier
$\Gamma_{G}$ by $w(G)$.

Given the epidemic size $n$, the inference on $p$ involves
evaluation of the likelihood function
$\mathcal{L}_n(p):=\mathbb{P}_p(|\mathcal{C}|=n)$ which can be
represented as follows:
$$
\mathcal{L}_n(p)=\sum\limits_{G\in\mathcal{G}_n}\mathbb{P}_p(G).
$$
As previously, under assumption of a uniform prior for $p$ its
posterior distribution $\pi(p\,|\, |\mathcal{C}|=n)$ is a mixture of
beta distributions:
\begin{equation}
\pi(p\,|\,
|\mathcal{C}|=n)\propto\sum\limits_{s,k,l}q(s,k,l)\thinspace
\text{Beta}\left(k+1,s-k+l+1\right), \label{eq:posteriorS2}
\end{equation}
where
$$q(s,k,l):=\#\{G\in\mathcal{G}_n\,|\, e(\satur{G})=s,e(G)=k,w(G)=l\}.$$

\begin{figure}[ptb]
\begin{center}
\includegraphics[width=3.2in]{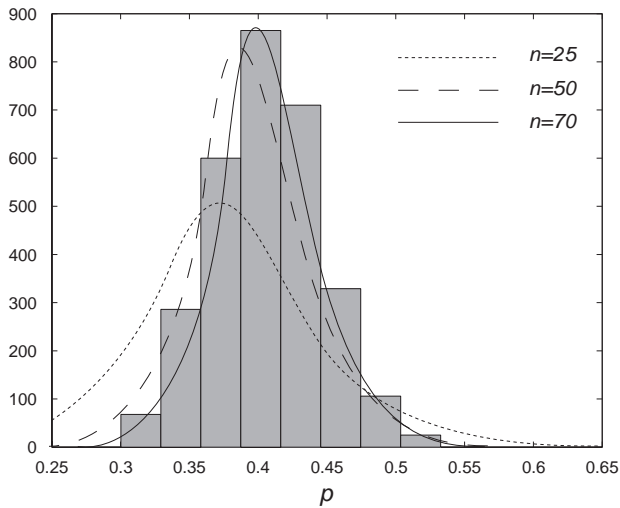}
\end{center}
\caption{Likelihood functions $\mathcal{L}_n(p)$ ($n=25,50,70$)
obtained using the McMC from Algorithm~\ref{alg:MCMC_S2}
and McMC sample histogram of $\mathcal{L}_n(p)$ for $n=70$.}%
\label{fig:likelihoods_for_connected_components}%
\end{figure}

This, again, represents a hard enumeration problem. However,
inference on $p$ can be made using the McMC technique.
Algorithm~\ref{alg:MCMC_S2} (Appendix~\ref{appendix:mcmc}) contains
a description of a Markov chain which serves the purpose of sampling
from the posterior distribution $\pi(p\,|\,|\mathcal{C}|=n)$, given
the prior distribution of $p$ is uniform on [0,1]. The chain
explores the joint space $(0,1)\times \mathcal{G}_n$ of possible
values for the percolation parameter $p$ and all possible connected
graphs on $n$ nodes, by deleting a vertex from a graph and adding a
vertex from its frontier.

Figure~\ref{fig:likelihoods_for_connected_components} depicts plots
of the likelihood function $\mathcal{L}_n$ when $n=25,\,50,\,70$ and
a histogram of a sample drawn from $\pi(p\,|\,|\mathcal{C}|=70)$.
The graphs of the likelihoods were obtained by smoothing the
histograms of corresponding samples generated by the McMC presented
in Algorithm~\ref{alg:MCMC_S2}. Note that, as might reasonably be
expected, the values of $p$ maximising the likelihood increase with
$n$.

Mixing properties of Markov chains described by
Algorithm~\ref{alg:MCMC_S2} are briefly discussed in
Appendix~\ref{appendix:Alg12_and_S12}. Two video examples of dynamic
graph updates illustrating Algorithms~\ref{alg:MCMC_S1} and
\ref{alg:MCMC_S2} can be found at the WEB address~\cite{BejanWEB}.

\subsection{Convergence analytical results for $\mathcal{S}2$}
Percolation exhibits a phenomenon of \textit{criticality}, this
being central the percolation theory: as $p$ increases, the sizes of
clusters (connected components) also increase, and there is a
critical value of $p_c$ at which there appears a cluster which
dominates the rest of the pattern. If $p<p_c$, then with probability
one all open clusters are finite, but when $p>p_c$ there is a single
infinite open cluster almost surely~\cite{Grimmett99}. Note, that
the critical percolation parameter $p_c$ depends on the dimension
$d$ of the lattice. Bond percolation on the square lattice seems to
be most studied to date of all percolation processes. The critical
probability $p_c$ in the case of a square lattice is $\frac{1}{2}$.
What follows, however, holds for any lattice $\mathbb{Z}^d$, $d\geq
3$.

\subsubsection{Asymptotic behaviour of the maximum likelihood estimates of~$p$}
Denote by $\mathcal{C}(x)$ the open cluster (connected component)
which contains the vertex $x$. Let us write
$\chi(p)=\mathbb{E}_p|\mathcal{C}|$ for the mean number of vertices
in the open cluster at the origin. Using the translation invariance
of the process on $\mathbb{Z}^d$, we have
$\chi(p)=\mathbb{E}_p|\mathcal{C}(x)|$ for all vertices $x$.
Percolation theory tells us that if $p<p_c$, then $\chi(p)<\infty$.
When $p>p_c$, then $\chi(p)=\infty$ and the function $\chi$ is not
of a much interest in this case. Instead, one studies the function
$\chi^{f}(p)=\mathbb{E}_p[|\mathcal{C}|:|\mathcal{C}|<\infty]$. The
function $\chi(p)$ ($\chi^{f}(p)$) monotonically increases to
infinity as $p\uparrow p_c$ ($p\downarrow p_c$). It is
known~\cite{Grimmett99} that there is no infinite open cluster when
$p=p_c$ for percolation on the square lattice $\mathbb{Z}^2$ and on
the lattice $\mathbb{Z}^d$, where $d\geq 19$; this is also believed
to be true for all other values of $d$. How likely it is to observe
an open cluster of size $n$ when $n$ is very large? What value of
$p$ should one suggest if one happened to observe a large epidemic
of size $n$?

Intuitively, it is unlikely that very large but finite epidemics may
be obtained for values of $p$ significantly different from $p_c$ (if
$p\ll p_c$, then it is unlikely that the epidemic will attain a very
large size; if $p\gg p_c$, then it is equally unlikely that, having
attained a sufficiently large size $n$, the epidemic will have
burned out). Intuition suggests therefore that the likelihood
function for $p$, given that the size $n$ of the connected component
containing the origin is growing, should be increasingly
concentrated around $p_c$.

Let $\mathcal{L}_n(p)=\mathbb{P}_p(|\mathcal{C}|=n)$ be the
probability that an open cluster $\mathcal{C}$ from a percolation
process with the edge density $p$ is of size $n$. (Equivalently,
assume that we observe an \textit{SIR} spread of infection via
nearest-neighbour interactions on $\mathbb{Z}^d$  and that the final
size of its outbreak is $n$; when $n$ is fixed the function
$\mathcal{L}_n(p)$ can be regarded as the likelihood function for
the percolation probability parameter $p$.)

Let $\hat{p}_n$ be the maximum likelihood estimate for $p$ derived
from $\mathcal{L}_n(p)$. For any $p$ define also
$L_n(p)=\mathcal{L}_n(p)/\mathcal{L}_n(p_c)$.

\begin{theorem}
The sequence of maximum likelihood estimates $\hat{p}_n$ for $p$
converges to the critical probability $p_c$. \label{thm:mle}
\end{theorem}

The proof of this theorem is based on the following lemma, together
with a generally accepted regularity condition.

\begin{lemma}
For any $p\in(0,1)$ different from $p_c$ the following holds:
$$
\liminf\limits_{n\to\infty}L_n(p)=0.
$$
Moreover, this convergence is uniform for any closed interval which
does not contain $p_c$.
\label{lem:p_c}
\end{lemma}

The proof of the lemma and the theorem can be found in
Appendix~\ref{appendix:proofs}.

As mentioned above, it is reasonable to expect the sequence of the
maximum likelihood estimates $\hat{p}_n$ to be monotonic (see
Figure~\ref{fig:likelihoods_for_connected_components}). This
observation gives rise to the following conjecture.
\begin{conjecture}
The sequence of $\{\hat{p}_n\}$ converges to $p_c$ monotonically
from the left. \label{conj:posteriors}
\end{conjecture}

Furthermore, we formulate the corresponding companion conjecture for
the sequence of posterior distribution, using the notion of a delta
sequence~\cite{Arfken1985}.
\begin{conjecture}
Provided $p\in\supp \pi(\cdot)$ the functional sequence
$\{\pi(p\,|\,n)\}_{n=1}^{\infty}$ is a delta sequence which
generates the delta function $\delta(p-p_c)$.
\label{conj:posterior_asympt}
\end{conjecture}
Thus, we believe that the limiting posterior distribution of the
percolation parameter is a one-point mass distribution at $p=p_c$,
or the Dirac delta function $\delta(p-p_c)$.

\subsubsection{Combinatorial characterisation of large percolation clusters on $\mathbb{L}^d$}
The theoretical results obtained and conjectured previously for
inference under scenario $\mathcal{S}2$ can be used to derive their
combinatorial analogues regarding the relative number of
realisations of the process with the cluster size $n$. Under
scenario $\mathcal{S}2$ the posterior distribution $\pi(p\,|\,
|\mathcal{C}|=n)$ can be seen as a mixture of beta distributions, as
in (\ref{eq:posteriorS2}). As in Section~\ref{sec:S2}, let
$q(s,k,l)$ be the number of all possible graphs $G$ which may
correspond to an open cluster $\mathcal{C}$ (the vertex set of $G$
and $\mathcal{C}$ coincide), and such that $k$ is the number of
edges in $G$, $s$ is the number of edges in the saturation of $G$,
and $l$ is the number of edges of $\Pi$ between the surface and the
frontier of $G$. Conjecture~\ref{conj:posterior_asympt} would
suggest that the number $q(s,k,l)$ of graphs $G$ corresponding to
open clusters $\mathcal{C}$ which could emerge as a result of the
percolation process with parameter $p_c$ and for which it holds that
\begin{equation}
\frac{k+1}{s+l+2}\approx p_c,
\end{equation}
is far greater than the number of all other graphs. This is so,
since the sequence of beta distributions
$\text{Beta}(\alpha_n,\beta_n)$ is a delta sequence generating the
delta function at $p_c$ if and only if
$\alpha_n/(\alpha_n+\beta_n)\to p_c$.

Thus, in percolation processes on $\mathbb{L}^d$ the number of
finite graphs corresponding to open clusters of size $n$ (where $n$
is large) that satisfy the condition
\begin{equation}
\frac{e(G)+1}{e(\satur{G})+w(G)+2}\approx p_c
\label{eq:approx_combinatorics_Ld}
\end{equation}
largely exceeds the number of all other connected components on $n$
nodes. In other words, a typical graph corresponding to an open
cluster of a large size in percolation process on $\mathbb{L}^d$ is
a connected subgraph of $\mathbb{L}^d$ characterised
by~(\ref{eq:approx_combinatorics_Ld}). In particular, when $d=2$:
\begin{equation}
e(G)-w(G)\approx e(\satur{G})-e(G);
\label{eq:approx_combinatorics_L2}
\end{equation}
that is the number of open edges in $G$ is approximately equal to
the total number of closed edges and edges between $G$ and its
frontier.

\subsubsection{Large percolation clusters as rare events}
When $n$ is large, the appearance of finite open clusters of size
$n$ is highly unlikely: the distribution of the cluster size
(hypothetically) decays as $n^{-1-1/\delta}$, $\delta>0$, when
$p=p_c$, and the decay is exponential (sub-exponential) when $p<p_c$
($p>p_c$). Large finite percolation clusters can therefore be viewed
as rare events. Since the state space of the McMC proposed for
inference on the percolation parameter $p$ under scenario
$\mathcal{S}2$ and described in Algorithm~\ref{alg:MCMC_S2} involves
the set of all open clusters on $n$ nodes, this algorithm can be
readily used in order to obtain realisations of these rare events.


\section{Bayesian optimal designs}
\subsection{Utility based optimal designs}
\label{subsection:utility_based_optimal_designs}
\subsubsection{General description}
The experimental design problem can be conveniently approached
within the Bayesian framework~\cite{Chaloner_Verdinelli1995}.
Suppose we study a stochastic process for which we formulate a model
$M$, characterised by a model parameter $\theta$. The model $M$ specifies a sampling density
 $f_d(y\,|\,\theta)$ for the
experimental outcome $y$ obtained from observation of the process under experimental
set-up $d$
given the value of the model parameter $\theta$. Our knowledge about
$\theta$ is described by a prior distribution $\pi(\theta)$.
Whenever the choice of set-up $d$ is under our control there
arises the question of selecting the optimal $d$ under which one
should observe the stochastic process. Such prescribed conditions
are referred to as a {\em design}, and the optimal design is found
under optimality criteria specifically tailored to
the context and the purpose of the experiment.

By employing a utility function $u(d,y,\theta)$ one can specify the
purpose of the experiment and measure the value of its outcome $y$
accordingly. The methodology of posing and solving utility-based
optimal design problems within the Bayesian paradigm became somewhat
standard~\cite{Muller1999,Cook&Gibson&Gilligan}. The design has to
be chosen before performing the experiment and one may choose to
maximise the expectation of the utility function $u(d,y,\theta)$
with respect to $\theta$ and $y$ \cite{Muller1999}:
\begin{equation}
d_{\text{max}}=\arg\max\limits_{d\in \mathcal{D}}U(d),
\label{eq:optimal_d}
\end{equation}
where
\begin{equation}
U(d)=\int\limits_{\Theta}\int\limits_{Y}
u(d,y,\theta)f_d(y\,|\,\theta)\pi(\theta)\, \text{d}\theta
\text{d}y. \label{eq:utility}
\end{equation}
Here $\mathcal{D}$ is the set of possible designs. The set of
possible outcomes $y$ of the experiment is denoted by $Y$. The
experiment is defined by a model $f_d(y\,|\,\theta)$, i.e.\ by the sampling density
 of $y$ conditional on $\theta$ for a given design~$d$.

When the purpose of the experiment is to infer the model parameter,
a sensible choice to measure the utility of the experiment under
design $d$ might be one that represents the information gained on
$\theta$ from the experiment. One of the standard choices is the
Kullback--Leibler (KL) divergence $D_{\text{KL}}$ between the prior
distribution of $\theta$, $\pi(\theta)$, and its posterior
distribution $\pi(\theta\,|\,y,d)$ obtained following the experiment
yielding data~$y$:
\begin{equation}
D_{\text{KL}}\{\pi(\theta\,|\,y,d)\parallel
\pi(\theta)\}=\int\limits_{\Theta}\log
\frac{\pi(\theta\,|\,y,d)}{\pi(\theta)}\pi(\theta\,|\,y,d)\,\text{d}\theta,
\label{eq:defining_KL_divergence}
\end{equation}
where $\pi(\theta\,|\,y,d)\propto f_d(y\,|\,\theta)\pi(\theta)$.

Introduced by Kullback and Leibler~\cite{Kulback&Leibler1951} this
information measure was justified by Lindley~\cite{Lindley56} and
Bernardo~\cite{Bernardo1979} and is related to the Shannon
information measure.

The KL divergence is a random variable, being a function of
the random outcome $y$. In order to find the
the design $d_{\text{max}}$ that is optimal with respect to this choice of utility
measure one should maximise the expected value of the KL divergence
over the set of observables $Y$ as a function of the design variate
$d$:
\begin{equation}
U_{\text{KL}}(d)=\mathbb{E}_Y[D_{\text{KL}}\{\pi(\theta\,|\,y,d)\parallel
\pi(\theta)\}], \label{eq:KL_expected_divergence}
\end{equation}
this being invariant under a reparametrisation of the model in terms
of $\theta$.

When the experimental motivation consists in increasing one's
knowledge on the model parameter the expected KL divergence
$U_{\text{KL}}(d)$ coincides with (\ref{eq:utility}) taking
$u(d,y,\theta)=\log \frac{\pi(\theta\,|\,y,d)}{\pi(\theta)}$. If,
however, the purpose of the experiment is to inform someone
who holds a different prior to our own and we wish to advise them on which design to use,
using our own `superior' knowledge
of the system under study, then the expected KL divergence should be calculated
in the form prescribed by (\ref{eq:KL_expected_divergence}) when the
integration over the space of observables $Y$ is to be carried out
using one's superior knowledge. We refer to these contrasting
experimental scenarios as \textit{progressive design} and
\textit{instructive design}~\footnote{In \cite{Cook&Gibson&Gilligan}
these experimental scenarios are called \textit{progressive} and
\textit{pedagogic} designs.}, respectively, and discuss them next
in more detail.

\subsubsection{Progressive design}
Under this scenario there is an experimenter $\mathscr{E}$ who holds
prior beliefs regarding the parameter, $\theta$, in the form of a prior
distribution $\pi(\theta)$ (perhaps obtained from earlier observation
of  the process) and whose goal is to design an optimal experiment in order to
increase this knowledge. Using KL divergence as the measure of information gain,
this experimenter should maximise the expected KL
divergence (\ref{eq:KL_expected_divergence}):

\begin{align}
U_{\text{KL}}(d)&=\mathbb{E}_Y\mathbb{E}_\Theta\left. \left[\log\frac{\pi(\theta\,|\,y,d)}{\pi(\theta)}\,\right|\,y\right] \nonumber\\
&=\int\limits_{\Theta}\int\limits_{Y}
\log\frac{\pi(\theta\,|\,y,d)}{\pi(\theta)}f_d(y,\theta)\,
\text{d}\theta \text{d}y,\nonumber
\end{align}
so that $U_{\text{KL}}(d)$ coincides with (\ref{eq:utility}) where
$u(d,y,\theta)=\log{\frac{\pi(\theta\,|\,y)}{\pi(\theta)}}$. Notice,
however, that the expected information gain $U_{\text{KL}}(d)$ can
be written as follows:

\begin{align}
U_{\text{KL}}(d)=&\int\limits_{Y}\int\limits_{\Theta}\log
\pi(\theta\,|\,y,d)f_d(y\,|\,\theta)\pi(\theta)\,\text{d}\theta\text{d}y\nonumber\\
&-\int\limits_{\Theta}\log \pi(\theta)
\pi(\theta)\,\text{d}\theta,\nonumber
\end{align}
where the second term is the Shannon entropy of $\pi(\theta)$, which
does not depend on $d$. Thus, in order to obtain the solution
$d_{\text{max}}$ in this case, it suffices to maximise the first
term only.

\subsubsection{Instructive design}
In contrast to the progressive design scenario, in the instructive
case there is an experimenter $\mathscr{E}$, holding a prior
$\pi(\theta)$, and a better (or differently) informed trainer (or
instructor) $\mathscr{T}$ whose knowledge about the model parameter
is summarised in a distribution $\pi^*(\theta)$. (Of particular
interest is a special case where the instructor knows the true value
of $\theta$.) The aim here is to maximise the change in
experimenter's belief from $\pi(\theta)$ to $\pi(\theta\,|\,y)$ by
designing an experiment using the superior knowledge
$\pi^*(\theta)$.

This latter optimisation problem can be formulated as follows:
\begin{align}
d^*_\text{max}&=\arg\max\limits_{d\in \mathcal{D}}U^*_{\text{KL}}(d),\label{eq:optimal_d_instructive}\\
U^*_{\text{KL}}(d)&=\int\limits_{Y}
D_{\text{KL}}\{\pi(\theta\,|\,y,d)\parallel \pi(\theta)\}f^*(y)\,
\text{d}y,\label{eq:utility_instructive}
\end{align}
where, as before, $\pi(\theta\,|\,y)$ is derived as in
(\ref{eq:defining_KL_divergence}), and $f^*(y)$ is the instructor's
prior predictive distribution for $y$:
$$
f^*(y)=\int\limits_{\Theta}f_d(y\,|\,\theta)\pi^*(\theta)\,\text{d}\theta.
$$
In particular, if the instructor $\mathscr{T}$ knows the exact value
of $\theta$, $\theta^*$, and hence $\pi^*(\theta)$ is the Dirac
function $\delta(\theta-\theta^*)$, then
$f^*(y)=f_d(y\,|\,\theta^*)$, so that
\begin{equation}
U^*_{\text{KL}}(d)=\int\limits_{\mathcal{Y}}
D_{\text{KL}}\{\pi(\theta\,|\,y,d)\parallel
\pi(\theta)\}f_d(y\,|\,\theta^*)\,\text{d}y.
\label{eq:instructive_delta}
\end{equation}

Note that the `instructive' expected utility $U^*_{\text{KL}}$ is
generally not symmetric with respect to the exchange of
$\pi(\theta)$ and $\pi^*(\theta)$: if the experimenter $\mathscr{E}$
and the instructor $\mathscr{T}$ change their roles, then the
corresponding optimal instructive designs are different, unless
their knowledge about $\theta$ is identical, that is
$\pi(\theta)=\pi^*(\theta)$ almost everywhere.

\subsection{Simulation-based evaluation of the expected utility}
\label{section:simulation based evaluation of expected utility}
Generally, the solution to the optimal design problems
(\ref{eq:optimal_d}-\ref{eq:utility}) and
(\ref{eq:optimal_d_instructive}-\ref{eq:utility_instructive}) cannot
be obtained analytically. This is mainly due to the following three
problems. First, the design space $\mathcal{D}$ can be complicated,
with many design variables, some of them having a continuous range of
values. Second, even if the design space has a simple structure, the
utility function may not be simple to evaluate, so the expected
utility $U(d)$ cannot be obtained explicitly. Finally, for
incomplete observations of highly non-linear stochastic processes
such as epidemic models, the likelihood is not usually available in
a closed form, and this results in computationally intensive
evaluations or sampling methods.

A review of analytical and approximate numerical solutions to
Bayesian optimal design problems for traditional experimental
design involving linear and non-linear models can be found in
\cite{Verdinelli1992,Chaloner_Verdinelli1995}.

M\"{u}ller~\cite{Muller1999} reviews simulation-based methods for
optimal design problems where the expected utility $U(d)$ is
evaluated by Monte-Carlo simulation. In its simplest form an
estimate $\widehat{U}$ of $U$ for any given design $d$ in the
progressive case is as follows:
\begin{equation}
\widehat{U}(d)=\frac{1}{M}\sum\limits_{i=1}^{M}u(d,\theta_i,y_i),
\label{eq:MCevaluation}
\end{equation}
where $\{(\theta_i,y_i),i=1,\ldots,M\}$ is a Monte-Carlo sample
generated values:
\begin{equation}
\theta_i\sim \pi(\theta), \thickspace y_i\sim p_d(y\,|\,\theta).
\end{equation}
The expected utility $U$ may, in particular, be based on the KL
divergence.

Analogously, the expected utility $U^*_{\text{KL}}$ under
the instructive scenario, when the instructor knows the true value of
the model parameter, $\theta^*$, can be evaluated using the
following scheme:

\begin{equation}
\widehat{U}^*_{\text{KL}}(d)=\frac{1}{M}\sum\limits_{i=1}^{M}\widehat{KL}(y^*_i,d),
\label{eq:*MCevaluation}
\end{equation}
where $y^*_i\sim f(y\,|\,\theta^*,d)$, $i=1,\ldots,M$, and
$\widehat{KL}(y_i,d)$ is an estimate of the KL divergence
$D_{\text{KL}}\{\pi(\theta\,|\,y^*,d)\parallel \pi(\theta)\}$ that
can be obtained via numerical integration of
$\log\frac{\pi(\theta\,|\,y,d)}{\pi(\theta)}$ with respect to the
posterior $\pi(\theta\,|\,y,d)$. The former function, in turn, may
need to be evaluated through simulation methods---perhaps using an
McMC scheme.

Evaluation of a continuous expected utility surface $U(d)$ by
(\ref{eq:MCevaluation}) or (\ref{eq:*MCevaluation}) can be achieved by
computing its values on a discretised grid of points and further
smoothing of the obtained set of values in order to approximate the
expected utility landscape. This, however, may be problematic when
$d\in\mathcal{D}$ is a multidimensional design parameter. An
alternative method is provided by the augmented probability
simulation approach which is studied in
\cite{Clyde&Mueller&Parmigiani1995,Bielza&Mueller&Rios-Insu}, and
reviewed in \cite{Muller1999}.

The augmented probability simulation approach assumes that
$u(d,\theta,y)$ is a non-negative bounded function. (This condition,
although not always automatically satisfied, can be easily achieved
by correspondingly modifying the utility function.) An artificial
density, proportional to $u(d,\theta,y)f_d(y\,|\,\theta)\pi(\theta)$
can then be defined for $(d,\theta,y)$ jointly~\cite{Muller1999}:
\begin{equation}
h(d,\theta,y)\propto u(d,\theta,y)f_d(y\,|\,\theta)\pi(\theta),
\end{equation}
so that its marginal in $d$ is proportional to the expected utility
$U(d)$. Sampling from $h(\cdot,\cdot,\cdot)$ can be used to obtain a
sample from its marginal using the Metropolis--Hastings McMC scheme
described in \cite{Muller1999}.

When using the augmented probability modelling approach one
approximates the optimal design by an empirical mode of the marginal
density of $d$ under the artificial distribution $h$, that is by the
mode of histogram formed from the first components of the set of
samples obtained. Notice that the evaluation of $u(d,\theta,y)$ at
each sampling step may in general itself involve an McMC sampling
from the posterior $\pi(\theta\,|\,y,d)$, so that the approach is
potentially very expensive computationally.

\subsection{Inner-outer plots}
We now apply the utility-based Bayesian approach introduced above in
order to design optimal experiments for our \textit{SIR} epidemic
model with incomplete observations of the kind described by
scenario~$\mathcal{S}1$. We do so by identifying a (finite) class of
designs (node configurations) which we call \textit{inner-outer
plots}. These are plots of bounded size which are obtained by
removing some nodes from the underlying grid. We refer to this
process as \textit{sparsification of the grid}.

\begin{figure}[ptb]
\begin{center}
\includegraphics[width=2.8in]{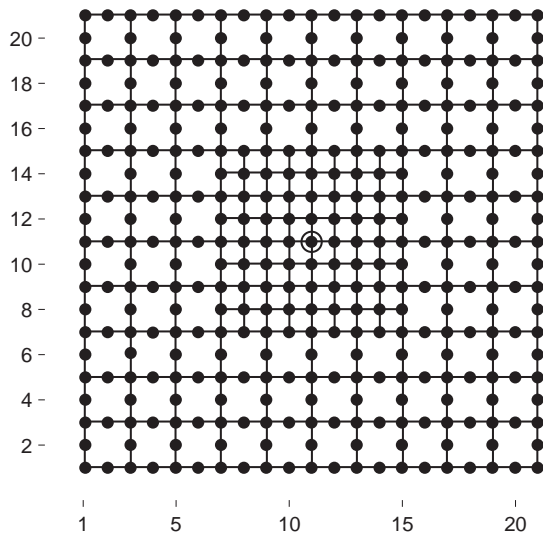}
\end{center}
\caption{An example of an inner-outer $(m,r)$-plot (nodes only) in
$\mathbb{Z}^2$. In this inner-outer plot $m=9$ and $r=3$. The plot
is bounded by an $N\times N$ square, where $N=21$ can be calculated using~(\ref{eq:m_on_N&r}).
Depicted is the saturation of an $(9,3)$-plot with respect to $\mathbb{L}^2$.}%
\label{fig:inner_outer_design}%
\end{figure}

A typical example of an inner-outer plot in $\mathbb{Z}^2$ is given
in Figure~\ref{fig:inner_outer_design}. The inner-outer plot
depicted consists of two parts---\textit{inner} and \textit{outer}
regions (hence the name of the design). Some sites are removed from
the outer region of the plot in a regular manner, while all sites of
the inner region are preserved. Our motivation in considering
sparsified grids is to show that intuitively obvious designs, such
as completely dense grids, can be improved. The justification of the
grid sparsification is that it impedes the ability of the epidemic
to spread, which makes possible better inference when $p$ is large,
while the more dense inner part of the design works better when $p$
is small. A formal description of inner-outer plots follows next.

Let us assume that $m$ is an odd positive integer and
$r\in\mathbb{N}\cup\{0\}$. An \textit{inner-outer $(m,r)$-plot
$\Pi^{(d)}_0(m,r)$ in $\mathbb{Z}^d$ with centre at the origin} is a
$d$-dimensional box $B_N^{(d)}$ with side-length $N=m+4r$ and some
vertices removed as follows:

\begin{equation}
\Pi_0(m,r):=
\begin{cases}
B_N^{(d)}, & r=0\\
B_N^{(d)}\setminus C_{N}^{(d)}(m,r), & r>0,
\end{cases}
\label{eq:inner-outer_definition}
\end{equation}
where $C_{N}^{(d)}(m,r)$ is the set
\begin{align}
\{&x\in B_N^{(d)}:\, \parallel x
\parallel_\infty=m+2j+1,j=0,\ldots,r-1\,\&\,\nonumber\\
&\&\parallel x\parallel_1\equiv0\bmod 2\}.\nonumber
\end{align}
Here $\parallel x \parallel_1=\sum\limits_{i=1}^d|x_i|$ and
$\parallel x
\parallel_\infty=\max (|x_1|,\ldots,|x_d|)$ for any
$x=(x_1,\ldots,x_d)\in\mathbb{Z}^2$ and the box $B_N^{(d)}$ is
defined as follows~\footnote{Note that $N$ is a positive integer
number since $m$ is odd.}:
\begin{align}
B_N^{(d)}:&=[-(N-1)/2,(N-1)/2]^d\nonumber\\
          &=\{x\in\mathbb{Z}^d:\,
\parallel x
\parallel_\infty\leq (N-1)/2\}.\nonumber
\end{align}
We call any plot that can be obtained by translating the plot
$\Pi^{(d)}_0(m,r)$ in $\mathbb{Z}^d$ an \textit{inner-outer
$(m,r)$-plot}, or simply \textit{an inner-outer plot}, and denote it
by $\Pi^{(d)}(m,r)$.

The total number of nodes contained in an $(m,r)$-plot can be
calculated by subtracting the total number of the nodes removed from
the outer plot (see (\ref{eq:inner-outer_definition})) as follows:
\begin{align}
T&=N(m,r)^2-4\sum\limits_{i=1}^{r}\left(\frac{m-1}{2}+2i-1\right)\nonumber\\
 &=(m+3r)^2+r(3r+2), \label{eq:Total_no_nodes_sparsified_plot}
\end{align}
where, as before,
\begin{equation}
N(m,r)=m+4r. \label{eq:m_on_N&r}
\end{equation}

The inner-outer $(m,r)$-plot depicted in
Figure~\ref{fig:inner_outer_design} is from $\mathbb{Z}^2$. The size
of the inner plot is $m\times m$, where $m=9$, and there are $r=3$
`circles' (with respect to the metric $\parallel \cdot
\parallel_\infty$) in the outer plot from which every second node is
removed. The size of the bounding box is $N\times N$, where
$N=m+4r=21$. The total number of nodes $T$ that this node
configuration contains can be calculated using
(\ref{eq:Total_no_nodes_sparsified_plot}) and is equal to $357$.

\subsection{Inference for \textit{SIR} epidemics on inner-outer plots}
We consider an \textit{SIR} epidemic with constant infectious times
which arises at the central site of an inner-outer plot
$\Pi=\Pi^{(d)}(m,r)$ from $\mathbb{Z}^d$ and evolves on that plot
according to the nearest-neighbour interaction rule, that is using
edges from $\mathbb{L}^d$ with endpoints from $\Pi$,  subject to
being restricted by the plot boundary. We consider this model in the
context of scenario $\mathcal{S}1$, that is when the only
information available about the outcome of the epidemic is its site
configuration $\mathcal{C}$. As before, this model is equivalent to
that of percolation on $\Pi^{(d)}(m,r)$.

\begin{figure*}[ptb]
\centering
\includegraphics[width=5.0in]{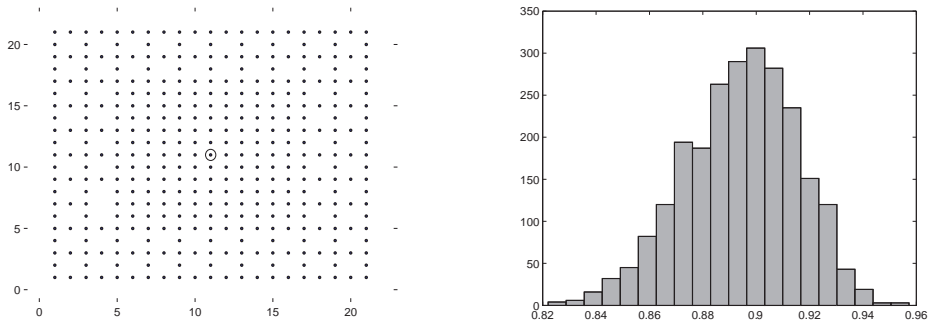}
\caption{Left: site configuration obtained on the inner-outer
$(13,2)$-plot with $p=0.9$.
Right: McMC histogram of an McMC sample for $p$ assuming a uniform prior for this parameter.}%
\label{fig:9025}%
\end{figure*}

\begin{figure*}[ptb]
\centering
\includegraphics[width=5.0in]{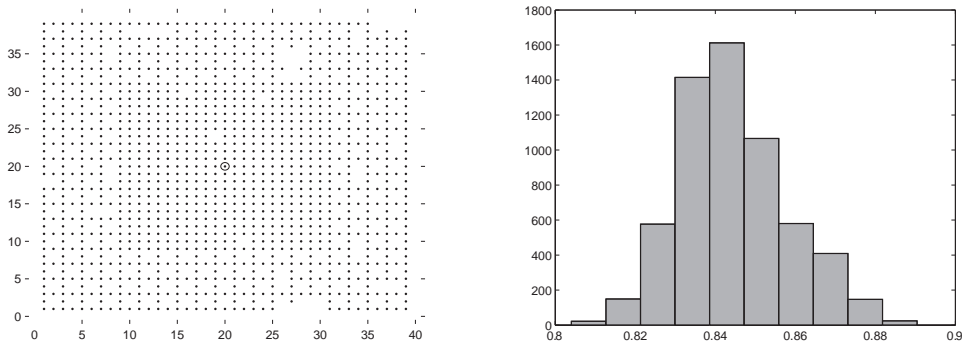}
\caption{Left: site configuration obtained on the inner-outer
$(23,4)$-plot with  $p=0.86$.
Right: histogram of an McMC sample for $p$ assuming a uniform prior for this parameter.}%
\label{fig:08616}%
\end{figure*}

We note that Algorithm~\ref{alg:MCMC_S1} and
Algorithm~\ref{alg:MCMC_S2} can be readily used for making inference
on the percolation probability parameter $p$ given the sites of an
open cluster $\mathcal{C}$, or merely its size in the case of
Algorithm~\ref{alg:MCMC_S2}, generated from a percolation process on
any locally finite graph~\footnote{By a locally finite graph we mean
a graph with no vertices of infinite degree.}.

Figure~\ref{fig:08616} shows the plot of a simulated open
percolation cluster $\mathcal{C}$ obtained on the inner-outer plot
$\Pi^{(2)}(13,2)$ with $p=0.9$ (left plot) and an estimate of the likelihood
formed from a histogram estimate of the
posterior density of $p$ obtained using the Markov chain described in
Algorithm~\ref{alg:MCMC_S1} with $\mathcal{C}$ as the input data, with a
uniform prior density for $p$.
Figure~\ref{fig:9025} shows similar plots for a realisation of the
percolation process on $\Pi^{(2)}(23,4)$ with $p=0.86$.

\subsection{Optimal design problem and design space}
We consider now an optimal design problem for the percolation
process on node sets of limited size. We adopt the utility-based
Bayesian approach formulated in
Section~\ref{subsection:utility_based_optimal_designs}. The choice
of the design space $\mathcal{D}$ can be made in a number of ways,
possibly reflecting such restrictions as limitations on the number
of experimental units or the dimensions of the experimental plot. In
the context of inner-outer plots the former restriction would mean
that $T$ from (\ref{eq:Total_no_nodes_sparsified_plot}) is bounded,
whereas the latter condition is equivalent to bounding the quantity
$N(m,r)$. A combination of these conditions or some other
information can be also taken into account when identifying the
design space. One advantage of grid-based plot designs is that they
can greatly reduce the dimensionality of the design space compared
with designs that allow units to be placed at arbitrary locations.

We assume now that $N$ is odd and fixed and that the design space has
the form
$$
\mathcal{D}=\{d\,|\, d=\Pi^{(2)}(m,r)\thickspace \&\thickspace
m+4r=N\}
$$
For example, if $N=19$ then, as can be derived easily from
(\ref{eq:m_on_N&r}), the design space consists of the following
designs:
\begin{align}
\mathcal{D}=\{&\Pi^{(2)}(19,0),\,\Pi^{(2)}(15,1),\,\Pi^{(2)}(11,2),\,\nonumber\\
              &\Pi^{(2)}(7,3),\,\Pi^{(2)}(3,4)\}.\nonumber
\end{align}
The set of observables $Y$ is the set of all connected components on
$d\in\mathcal{D}$ containing the central node.

\section{Practical implementation}
\label{subsection:practical_implementation} Given a finite design
space $\mathcal{D}$ it is fairly straightforward to solve the
optimal design problem for the percolation model on inner-outer plots
using the tools developed above. We now make some comments about
solving the problem under each of the two design scenarios.

\subsection{Progressive design: expected utility evaluation through augmented modelling}
Since $\mathcal{D}$ is finite we choose to identify the design that
maximises the expected utility function in the progressive case
using augmented modelling as described in
Section~\ref{section:simulation based evaluation of expected
utility}. Recall that this is based on an artificial distribution
$h(d,p,y)\propto u(d,p,y)f_d(y\,|\,p)\pi(p)$ from which samples are
generated using a Metropolis--Hastings sampler. The optimal design $d^*$
is identified then as a value of $d$ at which the marginal of $h$ is
maximised.

\subsection{Instructive design: Monte Carlo evaluation of the expected utility}
We treat the `instructive' case differently from that of the
`progressive' one because of the form of the expected utility under
this scenario. Recall that in the `instructive' case whenever the
instructor knows the true value $p^*$ of the model parameter $p$ one
can write the expected utility based on the Kullback--Leibler
divergence as follows (compare with (\ref{eq:instructive_delta})):
$$
U^*_{\text{KL}}(d)=\sum\limits_{y\in\mathcal{Y}}
f_d(y\,|\,p^*)\int\limits_0^1\log\frac{\pi(p\,|\,y,d)}{\pi(p)}\pi(p\,|\,y,d)\,\text{d}p,
$$
where $f_d(y\,|\,p^*)$ is the likelihood function evaluated at the
true value of the model parameter $p^*$ given the open cluster $y$
at the centre of the inner-outer plot $d$.

It is clear that to evaluate the expected utility
$U^*_{\text{KL}}(d)$ via standard Monte Carlo simulation, then a
Markov chain has to be run to evaluate the posterior
$\pi(p\,|\,y,d)$ each time we sample for a new observation (an open
cluster) $y$ and also the potentially time-consuming integration has
to be done with respect to the model parameter~$p$. This integration
can be implemented in the following way. Since we can sample from
the posterior $\pi(p\,|\,y,d)$ via McMC
(Algorithm~\ref{alg:MCMC_S1}) for any given open cluster $y$, we do
so and then fit the beta distribution (or some other distribution)
to the McMC sample obtained in order to perform integration
numerically in a more efficient, albeit, approximate manner.

Thus, the expected utility evaluation scheme for inner-outer plots
in the `instructive' case and scenario $\mathcal{S}1$ can be
described as follows.

For each inner-outer plot $d\in\mathcal{D}$ we do the following:
\begin{enumerate}
\item generate a random sample of $M$ independent connected clusters $\{y_i\}_{i=1}^{M}$
on $\Pi(m,r)$: $y_i\sim f(y\,|\,p^*,d)$;
\item perform $M$ McMC's in order to obtain $M$ independent samples for the posterior distribution
$\pi(p\,|\,y,d)$;
\item fit a Beta distribution to each of the samples obtained; refer to the fitted distributions as
$\pi(p\,|\,y_i,d)$;
\item evaluate numerically the integrals
$$I_i:=\int_{0}^{1}\log\frac{\pi(p\,|\,y_i,d)}{\pi(p)}\pi(p\,|\,y_i,d)\,\text{d}p;$$
\item estimate the expected utility: $\bar{U}_M=\frac{1}{M}\sum\limits_{i=1}^{M}I_i.$
\end{enumerate}

It is natural to ask how well the true posteriors can be fitted by a
Beta distribution (step~(iii) of the above scheme). Experience
suggests that such a fit never affects the outcome of the analysis
on the qualitative level unless the prior distribution has more than
one local mode or its support is smaller than the entire interval
$(0,1)$. Recall that the purpose of this step is to make evaluation
of the integrals $I_i$ easier and faster, and hence the family of
Beta distributions is only one possible choice. For instance, if the
prior distribution $\pi(p)$ is uniform on $(0,1)$, then the
integrals $I_i$ represent the entropies of the fitted distributions.
In the case when the fitting is done by beta distributions, each of
these integrals can be quickly calculated using the following
analytical formula for the entropy of the beta distribution
$\text{Beta}(\alpha,\beta)$:
\begin{align}
\ent{\text{Beta}(\alpha,\beta)}=&\log
B(\alpha,\beta)+(\alpha+\beta-2)\psi(\alpha+\beta)\nonumber\\
&-(\alpha-1)\psi(\alpha)-(\beta-1)\psi(\beta),\nonumber
\end{align}
where $\psi$ is the \textit{digamma function}, $\psi(z)=\Gamma
'(x)/\Gamma(z)$. Other choices of prior distribution may
suggest alternative and more appropriate fitting distributions to facilitate
the calculation of the integrals $I_i$,
$i=1,\ldots,M$.

\subsection{Example}
\begin{figure}[ptb]
\begin{center}
\includegraphics[width=3.5in]{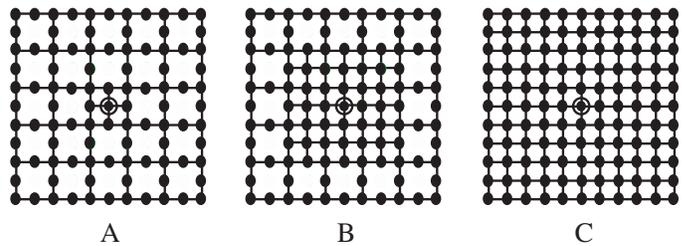}
\end{center}
\caption{\textit{Inner-outer} design plots $A$, $B$, and $C$ form the design space $\mathcal{D}=\{A,B,C\}$.}%
\label{fig:inner_outer_sample}%
\end{figure}

In our example we consider all inner-outer plots in $\mathbb{L}^2$
whose sizes do not exceed $N=11$. There are only three such plots:
$\Pi^{(2)}(3,2)$, $\Pi^{(2)}(7,1)$, and $\Pi^{(2)}(11,0)$. For ease
of reference we mark them $\text{A}$, $\text{B}$, and $\text{C}$
respectively (as depicted in Figure~\ref{fig:inner_outer_sample}).
Thus, the design space $\mathcal{D}=\{\text{A},\text{B},\text{C}\}$
consists of three designs, among which $\text{A}$ is the most
sparsified plot whereas no nodes are removed from $\text{C}$ at all.

\begin{figure*}[ptb]
\begin{center}
\includegraphics[width=5.4in]{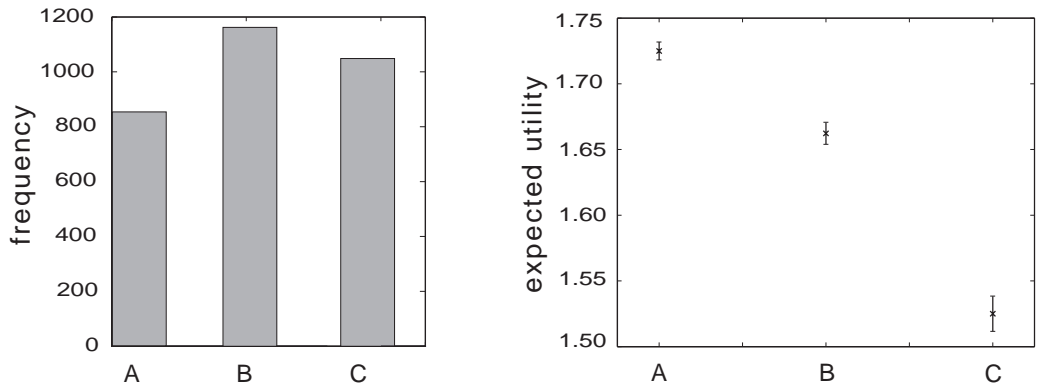}
\end{center}
\caption{Left: sample histogram for the marginal of $h(d,p,y)$ in
$d$, $d\in\{\text{A},\text{B},\text{C}\}$, under progressive design
and $\pi(p)\sim\text{U}(0,1)$. Right: evaluated expected utility
under instructive design with $\pi^*(p)\equiv\delta(p-0.9)$ and 95\%
credibility intervals ($M=1500$) for the plots $\text{A}$,
$\text{B}$, and $\text{C}$,
under instructive design.}%
\label{fig:inner_outer_hist}%
\end{figure*}

Figure~\ref{fig:inner_outer_hist} represents graphically the results
of the comparison of designs from $\mathcal{D}$ under both
`progressive' and `instructive' scenarios when the prior distribution
$\pi(p)$ is uniform on the interval $(0,1)$. The left panel of the
figure corresponds to the former scenario and depicts a histogram of
a sample corresponding to the marginal of the artificial augmenting
distribution $h(d,p,y)\propto u(d,p,y)f(y\,|\,p,d)\pi(p)$ in
$d\in\mathcal{D}$. The right panel corresponds to the latter
scenario and shows the Monte Carlo estimated values of the expected
utilities and 95\% credibility intervals for each of the three
considered designs ($M=1500$, see (\ref{eq:MCevaluation}) in
Section~\ref{section:simulation based evaluation of expected
utility}) assuming that the instructor knows $p$ precisely
so that $\pi^*(p)=\delta(p-0.9)$.

The plots from Figure~\ref{fig:inner_outer_hist} indicate that the
solutions to the optimal design problem under the two scenarios are
different from each other. The `moderately sparsified'
plot~$\text{B}$ maximises the expected utility in the progressive
case, that is in the case when the experiment is designed by a
single experimenter. If, however, it is the instructor who knows the
true value of the model parameter ($p=0.9$) and wishes to choose the
best inner-outer plot from the set $\mathcal{D}$ for a more
uninformed experimenter to use (instructive scenario), then the
optimal plot is the `mostly sparsified' inner-outer plot $\text{A}$.
Notably, the `mostly dense' plot $\text{C}$ would be the worst
choice in the instructive case, whereas it outperforms the `mostly
sparsified' plot $\text{A}$ but is worse than the `moderately
sparsified' plot $\text{B}$ in the progressive case.  These results
demonstrate the moderately counter-intuitive finding that the
optimal design is not necessarily that with the maximal number of
experimental units and that a 'less-is-more' principle may be in
operation.  On another level, this finding may not be entirely
unpredictable in the instructive case.  The optimal design's
combination of a dense centre and sparsified outer region may help
to increase the probability that the experimental epidemic succeeds
in establishing itself but does not lead to saturation of the
lattice before it dies out.  This in turn may represent an outcome
that yields much information on the unknown~$p$.

\section{Concluding comments}

In this paper we have considered a number of question relating to
the observation of percolation processes on finite lattices.  We
have demonstrated the value of McMC techniques for parameter
estimation under different sampling regimes and shown how Bayesian
experimental design techniques can be used to identify the optimal
lattices on which to carry out experiments.

We introduced inner-outer plot designs by sparsifying the underlying
grid. Allowing one to control excessive connectivity in outer
regions of experimental plots, these designs help to separate
different values of the model parameter $p$ when $p\gg p_c$, while
giving the epidemic a chance to establish in inner regions when
$p\ll p_c$. Inner-outer plots can be further generalised by
considering more than two levels of grid sparsification, according
to the principle used in our work: the farther the region is from
the centre of the plot the more intensive sparsification it
receives. To an extent, this process can be viewed as switching to
similarly shaped lattices with percolation parameter taking its
values from the set $\{p,p^2,p^3,\ldots\}$ in decreasing fashion.

Although we have worked largely within framework of percolation
theory, the connection between percolation and spatio-temporal
epidemic models with nearest-neighbour interactions mean that the
results may have significant implications for and application to the
study of epidemic systems.

\begin{acknowledgments}
We thank Dr~Alex Cook for useful discussions at early stages of this
work. A.~Iu.~B. acknowledges James Watt scholarship, Overseas
Research Student Awards Scheme support and funding provided through
the TIME project (UK EPSRC grant EP/C547632/1).
\end{acknowledgments}

\appendix
\section{Algorithms and proofs}
\subsection{McMC algorithms for the scenarios~$\mathcal{S}1$ and $\mathcal{S}2$}
\label{appendix:mcmc}
An McMC algorithm for making inference on the percolation
probability $p$ under scenario $\mathcal{S}1$ is given in
Algorithm~\ref{alg:MCMC_S1}. The proposed McMC is irreducible by
construction: there is a positive probability for the chain to
switch between any two connected graphs from
$\mathcal{G}(\mathcal{C})$, since any two such graphs share a common
vertex set and differ only by a finite number of edges.

\begin{algorithm}[ptb]
\caption{Markov Chain Monte Carlo: scenario $\mathcal{S}1$}
\begin{algorithmic}[1]
\REQUIRE an open cluster $\mathcal{C}$;


\STATE take an initial value $p_0$ arbitrary from $(0,1)$;

\STATE $t:=0$ $X_t:=(p_t,\satur{\mathcal{C}})$;

\REPEAT

\STATE \underline{\textit{Gibbs sampler steps:}}

\STATE
$p_{2t+1}\sim\text{Beta}\left(e(G_{2t})+1,e(\satur{\mathcal{C}})-e(G_{2t})+w_\mathcal{C}+1\right)$;

\STATE $X_{2t+1}:=(p_{2t+1},G_{2t})$;

\STATE \underline{\textit{Metropolis--Hastings sampler steps:}}

\STATE choose an edge $e$ uniformly at random from
$\satur{\mathcal{C}}$;

\IF{$e\in G_{2t}$}

\STATE $U\sim\text{Uniform}[0,1]$;

\IF{
$U\leq\min(1,\frac{1-p_{2t+1}}{p_{2t+1}}\indicator{G_{2t+1}\setminus
\{e\} \text{ is connected}})$}

\STATE $X_{2t+2}:=(p_{2t+1},G_{2t+1}\setminus\{e\})$;

\ELSE \STATE $X_{2t+2}:=(p_{2t+1},G_{2t+1})$;

\ENDIF

\ELSE \STATE $U\sim\text{Uniform}[0,1]$;

\IF {$U\leq \min(1,\frac{p_{2t+1}}{1-p_{2t+1}})$} \STATE
$X_{2t+2}:=(p_{2t+1},G_{2t+1}\cup\{e\})$; \ELSE \STATE
$X_{2t+2}:=(p_{2t+1},G_{2t+1})$; \ENDIF

\ENDIF

\STATE $t:=t+2$;

\UNTIL{we judge that the chain has converged and a sample of
sufficient size is obtained}
\end{algorithmic}
\label{alg:MCMC_S1}
\end{algorithm}


Algorithm~\ref{alg:MCMC_S2} describes an McMC algorithm for making
inference on $p$ under scenario $\mathcal{S}2$. This algorithm,
similarly to Algorithm~\ref{alg:MCMC_S1}, is a combination of Gibbs
and Metropolis--Hastings steps. The chain's stationary distribution
$f(p,G)$ specified a marginal density for $p$ which coincides with the posterior
distribution $\pi(p\,|\,|\mathcal{C}|=n)$.

\begin{algorithm}[ptb]
\caption{Markov Chain Monte Carlo: scenario $\mathcal{S}2$}
\begin{algorithmic}[1]
\REQUIRE the value of $n$.


\STATE take a value $p_0$ arbitrary from $(0,1)$ and a graph $G_0$
arbitrary from $\mathcal{G}_n$;

\STATE $t:=0$ $X_t:=(p_t,G_t)$;

\REPEAT

\STATE \verb+move_on+:=1;

\STATE \underline{\textit{Gibbs sampler steps:}}

\STATE
$p_{2t+1}\sim\text{Beta}\left(e(G_{2t})+1,e(\satur{G_{2t}})-e(G_{2t})+w(G_{2t})+1\right)$;

\STATE $X_{2t+1}:=(p_2t+1,G_{2t})$;

\STATE \underline{\textit{Metropolis--Hastings sampler steps:}}

\STATE choose a vertex $u$ uniformly at random from $G_{2t+1}$ and
choose a vertex $v$ uniformly at random from $\Gamma_{G_{2t+1}}$.
Derive a graph $\tilde{G}$ from $G_{2t+1}$ by deleting all edges
which adjoin $u$ (in $G_{2t+1}$) and adding the edges that connect
$v$ with vertices of the graph $G_{2t+1}\setminus\{u\}$ in $\Pi$,
each independently with probability $p_{2t+1}$ (conditioning on the
event that at least one edge is added).

\IF{$\tilde{G}$ is disconnected} \STATE
$X_{2t+2}:=(p_{2t+1},G_{2t+1})$; \verb+move_on+:=0 \ENDIF

\IF{$\texttt{move\textunderscore on}$}

\STATE $U\sim\text{Uniform}(0,1)$;

\STATE $\tilde{d}(v):=\#\{e\,|\,e=(v,z)\thinspace \exists z\in
G_{2t+1}\setminus\{u\}\}$;

\STATE $\tilde{d}(u):=\#\{e\,|\,e=(u,z)\thinspace \exists z\in
\tilde{G}\setminus\{v\}\}$;

\STATE $\nu(u):=\#\{x\,|\, x\in\Gamma_{G_{2t+1}} \,\&\,
(u,x)\in\Pi\}$;

\STATE $\nu(v):=\#\{x\,|\, x\in\Gamma_{\tilde{G}} \,\&\,
(v,x)\in\Pi\}$;

\STATE $\kappa:=\tilde{d}(u)-\tilde{d}(v)+\nu(v)-\nu(u)$;
$U\sim\text{Uniform}(0,1)$;

\IF{$U\leq
\min\left(1,\frac{|\Gamma_{G_{2t+1}}|}{|\Gamma_{\tilde{G}}|}
\frac{1-(1-p_{2t+1})^{\tilde{d}(v)}}{1-(1-p_{2t+1})^{\tilde{d}(u)}}(1-p_{2t+1})^{\kappa}\right)$}

\STATE $X_{2t+2}:=(p_{2t+1},\tilde{G})$;

\ELSE \STATE $X_{2t+2}:=(p_{2t+1},G_{2t+1})$;

\ENDIF

\ENDIF

\STATE $t:=t+2$;

\UNTIL{we judge that the chain has converged and a sample of
sufficient size is obtained}
\end{algorithmic}
\label{alg:MCMC_S2}
\end{algorithm}

We give now explicit expressions for the proposal probabilities used
in the Metropolis--Hastings steps  within this algorithm. Assume that the
current graph within the Metropolis--Hastings step is $G$ and a
graph $\tilde{G}$ is proposed, the latter being obtained from the
former by deleting a vertex $u$ with all edges adjoining it and
inserting a vertex $v$, including each potential edge to $v$ independently
with probability $p$, this probability having been drawn in the preceding Gibbs
step. We assume that at least one such edge is inserted and denote
the number of deleted and added edges by $d(u)$ and $d(v)$
respectively. Then,
\begin{equation}
q(G,\tilde{G})=\frac{1}{n}\frac{1}{|\Gamma_{G}|}
\frac{p^{d(v)}(1-p)^{\tilde{d}(v)-d(v)}}{1-(1-p)^{\tilde{d}(v)}}\indicator{\tilde{G}\text{
is connected}},
\end{equation}
and similarly,
\begin{equation}
q(\tilde{G},G)=\frac{1}{n}\frac{1}{|\Gamma_{\tilde{G}}|}\frac{p^{d(u)}(1-p)^{\tilde{d}(u)-d(u)}}{1-(1-p)^{\tilde{d}(u)}}
\indicator{G\text{ is connected}}.
\end{equation}

Clearly,
\begin{align}
\mathbb{P}_p(G)&\propto p^{d(u)}(1-p)^{\tilde{d}(v)}(1-p)^{\tilde{d}(u)-d(u)+\nu(u)}\nonumber\\
\mathbb{P}_p(\tilde{G})&\propto
p^{d(v)}(1-p)^{\tilde{d}(v)}(1-p)^{\tilde{d}(v)-d(v)+\nu(v)},\nonumber
\end{align}
so that the acceptance probability at the Metropolis--Hastings step,
$\alpha$, is as follows:
\begin{align}
\alpha&=\min
\left(1,\frac{q(\tilde{G},G)}{q(G,\tilde{G})}\frac{\mathbb{P}_p(\tilde{G})}{\mathbb{P}_p(G)}\right)\nonumber\\
&=\min
\left(1,\frac{|\Gamma_{G}|}{|\Gamma_{\tilde{G}}|}\frac{(1-p)^{\tilde{d}(u)+\nu(v)}}{(1-p)^{\tilde{d}(v)+\nu(u)}}
\frac{1-(1-p)^{\tilde{d}(v)}}{1-(1-p)^{\tilde{d}(u)}}\right)\nonumber\\
&=\min,
\left(1,\frac{|\Gamma_{G}|}{|\Gamma_{\tilde{G}}|}\frac{1-(1-p)^{\tilde{d}(v)}}{1-(1-p)^{\tilde{d}(u)}}(1-p)^\kappa\right),\nonumber
\end{align}
where, as it was introduced in the description of
Algorithm~\ref{alg:MCMC_S2},
$$\kappa=\tilde{d}(u)-\tilde{d}(v)+\nu(v)-\nu(u).$$

We claim that the constructed chain is irreducible. In terms of graph theory
this means that the chain can proceed from each connected graph
on $n$ vertices, including the origin, to any other connected graph
including the origin of the same size on the underlying lattice. We show
the irreducibility of the proposed McMC by constructing a sequence
of steps in which any graph of $\mathcal{G}_n$ is transformed to a
so called \textit{line-skeleton} graph on $n$ vertices. By such a
graph we mean any tree (a graph with no cycles) containing the
origin and having $n$ vertices so that only two of
these vertices have degree one. Select one such
line-skeleton and denote it by $S$.

Consider a graph $G$ from $\mathcal{G}_n$ and denote the length of
the shortest open path from $x\in G$ to $S$ by $\delta(x,S)$
(chemical distance). Each vertex $x$ from $G$ receives a well
defined finite weight $\delta(x,S)$, since the graph $G$ is
connected and finite. By using the description of our Markov chain
we can delete any vertex from our current graph for which
$\delta(x,S)$ is maximal and add to this graph a vertex from the
chosen line skeleton $S$ without making the graph disconnected or
containing cycles until the maximum value of $d(x,S)$ is zero---in
this case all vertices are forming the line skeleton. Since this
procedure can be reversed it follows that $\mathcal{G}_n$ in the
described Markov chain is indeed a communicating class, and hence
the chain is irreducible.

\subsection{Lemma and Theorem}
\label{appendix:proofs}
\begin{proof} (Lemma~\ref{lem:p_c})
We consider the following two cases~\cite{Grimmett99}:
\begin{enumerate}
\item \emph{Subcritical case} $p<p_c$. In this case the cluster size distribution $\mathcal{L}_n(p)$ decays
exponentially, i.e.\
\begin{equation}
\exists\, \lambda(p)>0: \thickspace \mathcal{L}_n(p)\leq
e^{-n\lambda(p)}\thickspace \forall n\geq 1. \label{eq:*}
\end{equation}
However, it is also known that at $p=p_c$ the cluster size
distribution, while almost surely finite, nevertheless has an
infinite mean. This implies that, for any $\lambda$ such that
$0<\lambda<\lambda_p$, we have
$\limsup\limits_{n\to\infty}e^{\lambda n}\mathcal{L}_n(p_c)=\infty$,
which when combined with (\ref{eq:*}), gives
$\liminf\limits_{n\to\infty} L_n(p)=0$, as required. Uniformity of
convergence on any closed interval contained within $(0,p_c)$
follows since $\lambda$ above may be chosen so that $\lambda_p$ is
uniformly bounded away from $\lambda$ for $p$ belonging to such an
interval.

\item \emph{Supercritical case} $p>p_c$. The argument here is essentially the same. In this case the decay is sub-exponential:
\begin{equation}
\exists\, \eta(p)>0:\thickspace \thickspace \mathcal{L}_n(p)\leq
e^{-\eta(p)n^{(d-1)/d}}\thickspace \forall n\geq 1. \label{eq:**}
\end{equation}
The infinite mean of the cluster size at $p=p_c$ also implies that
for any $\eta$ such that $0<\eta<\eta_p$, we have
$\limsup\limits_{n\to\infty}e^{\eta
n^{(d-1)/d}}\mathcal{L}_n(p_c)=\infty$, which, when combined with
(\ref{eq:**}), again gives $\liminf_{n\to\infty} L_n(p)=0$, as
required. Uniformity of convergence on any closed interval contained
within $(p_c,1)$ follows as before, and, when taken with the earlier
result, gives the uniformity assertion of the lemma.
\end{enumerate}
\end{proof}

\begin{proof} (Theorem~\ref{thm:mle})
We assume only that $\lim_{n\to\infty}L_n(p)$ exists. Then, by
Lemma~\ref{lem:p_c} this limit is equal to zero for all $p\ne p_c$.
(The existence of the limit would follow, for example, from the
widely believed, but not rigorously proved, result that for $p=p_c$
the distribution of the size of the open cluster size containing the
origin has a power-law tail---see Chapter~9 of \cite{Grimmett99}.)
Then the result of Lemma~\ref{lem:p_c} can be reformulated in
$\varepsilon$-terms as follows: $\forall \varepsilon>0\thickspace
\exists\, N(\varepsilon,\gamma)>0$, such that
\begin{equation}
\mathcal{L}_n(p)<\varepsilon\mathcal{L}_n(p_c)\, \forall
n>N(\varepsilon,\gamma)\, \forall p\in [\alpha,p_c-\gamma]\cup
[p_c+\gamma,\beta), \label{eq:L<epsilonL}
\end{equation}
for any $\alpha$ and $\beta$, such that
$$\Delta=[\alpha,p_c-\gamma]\cup
[p_c+\gamma,\beta],\,\Delta\subset(0,1).$$

The quantity $\hat{p}_n$, being the maximum likelihood estimate for
$p$, is the mode of $\mathcal{L}_n(p)$:
$$
\hat{p}_n:=\text{arg}\max\limits_{p\in(0,1)}\mathcal{L}_n(p),
$$
that is
\begin{equation}
\mathcal{L}_n(\hat{p}_n)\geq\mathcal{L}_n(p)\thickspace\forall
p\in(0,1). \label{eq:mle}
\end{equation}

Consider the sequence of mle's $\{\hat{p}_n\}_{n=1}^{\infty}$. We
now prove that this sequence converges to $p_c$. Suppose,
conversely, this is not the case:
$$
\exists\, \zeta\in(0,1)\thinspace \forall M>0\thinspace \exists\,
n>M:\thinspace |\hat{p}_n-p_c|>\zeta.
$$
Take $M(\zeta)=N(\zeta,\zeta/2)$, then $\exists\,
n>M(\zeta):\thinspace |\hat{p}_n-p_c|>\zeta$, and, thus,
$p_c\neq\hat{p}_n$. At the same time (when $\varepsilon=\zeta$) the
following holds by (\ref{eq:L<epsilonL}):
$$
\mathcal{L}_n(\hat{p}_n)<\zeta\mathcal{L}_n(p_c)<\mathcal{L}_n(p_c),
$$
thereby contradicting (\ref{eq:mle}). Hence,
$\hat{p}_n\rightarrow p_c$, $n\rightarrow\infty$.
\end{proof}

\section{Mixing properties of chains from the proposed algorithms}
\label{appendix:Alg12_and_S12} 
\begin{figure}[ptb]
\centering
\includegraphics[width=3.4in]{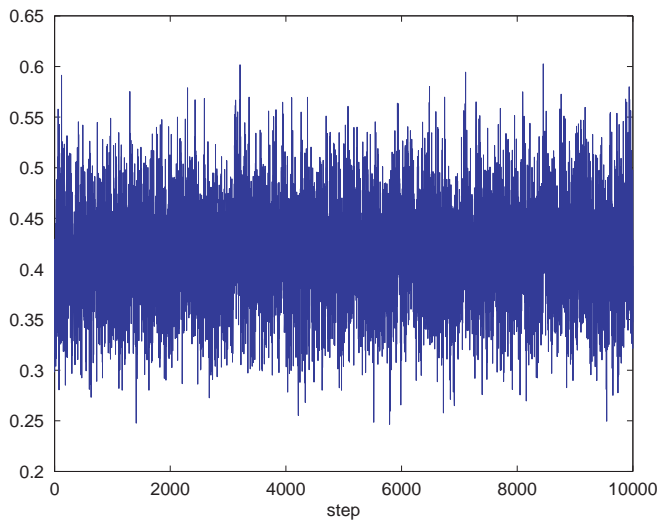}
\caption{Trace plot for McMC sampling resulted in the histogram from
Figure~\ref{fig:post_n_likelihood} for the cluster $\mathcal{C}$ in
Figure~\ref{fig:outbreak}.} \label{fig:traces}
\end{figure}

\textbf{Scenario $\mathcal{S}1$ and
Algorithm~\ref{alg:MCMC_S1}.} Figure~\ref{fig:traces} displays a
trace plot of updates in $p$ for inference made for the site
configurations from Figure~\ref{fig:outbreak}
(Algorithm~\ref{alg:MCMC_S1} was applied) and presented in
Figure~\ref{fig:post_n_likelihood}. The trace plot indicates that
the mixing properties of the chain are rather satisfactory. The
Matlab-based simulation used required 15 seconds on Intel(R)
Core(TM)2 Duo CPU~2.26GHz to obtain a series of chain updates of the
length~$10^4$.

Experiments with larger configurations suggest that the burn-in
time of the constructed chain may vary considerably with the
size of the site configuration and its potential connectivity
properties as well as on the choice of the initial graph $G_0$ (in
our experiments we worked with fully saturated graphs and with
connected components derived from those by random sparsification of
edges). However, in practice, the
chain described by Algorithm~\ref{alg:MCMC_S1} typically converges rapidly: both the
connected component updates and the percolation parameter updates
can be efficiently realised, allowing one to generate a suitable sample
from the posterior distribution $\pi(p\,|\,\mathcal{C})$ within practically useful
timescales. Updates of the connected component, that is deletion and
insertion of edges while preserving the \textit{connectedness} of
the underlying graph, can be further improved using dynamic graph
algorithms~\cite{Zaroliagis2002}.

\begin{figure*}[ptb]
\centering
\includegraphics[width=5.0in]{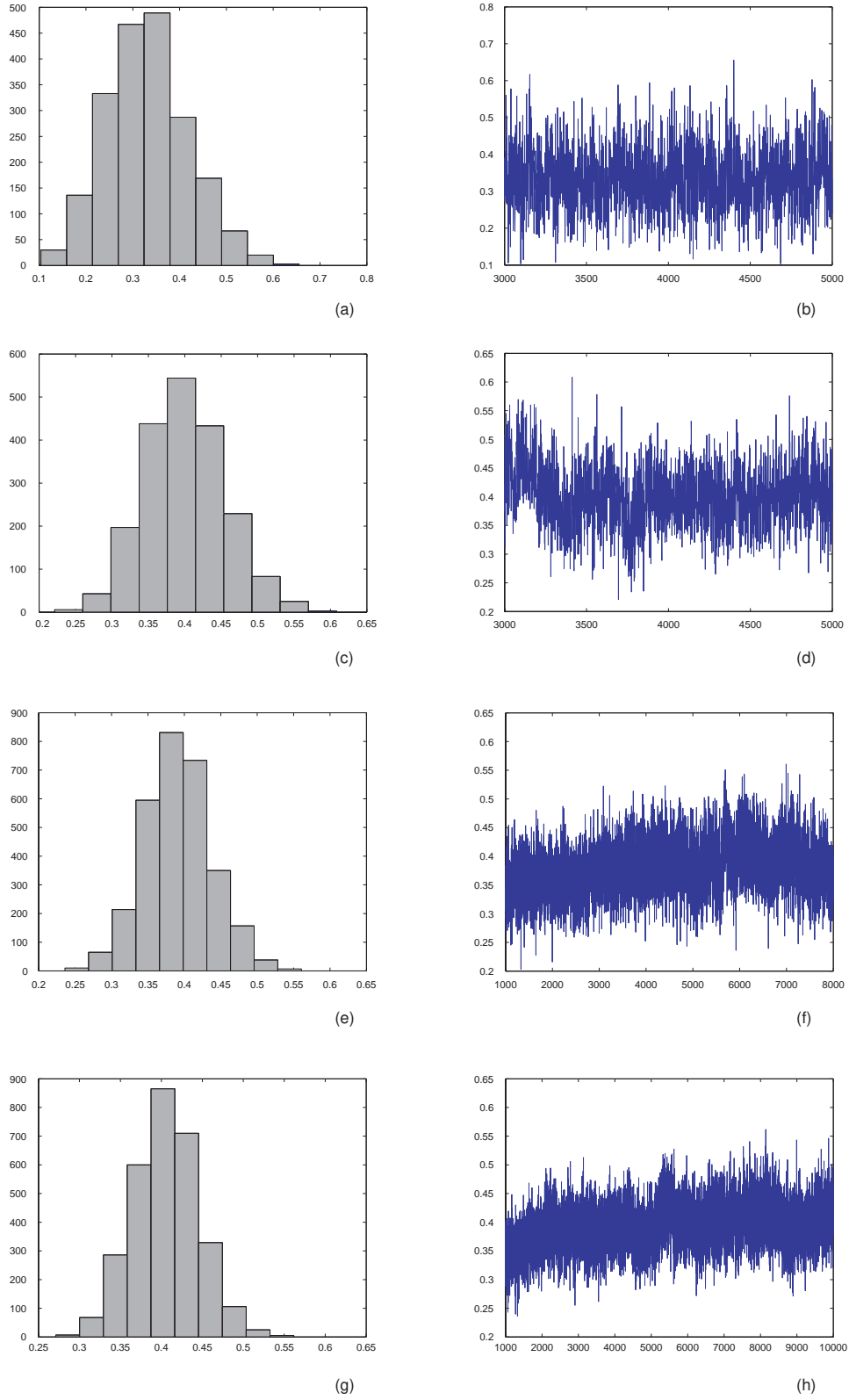}
\caption{Inference on the percolation parameter using McMC described
in Algorithm~\ref{alg:MCMC_S2}: histograms of obtained samples and
trace plots for (a,b) $n=10$; (c,d) $n=35$; (e,f) $n=50$; (g,h)
$n=70$.}
\label{fig:S2_hist_n_mixing}%
\end{figure*}

\textbf{Scenario $\mathcal{S}2$ and Algorithm~\ref{alg:MCMC_S2}.}
Figure~\ref{fig:S2_hist_n_mixing} shows histograms of samples from
the distribution $\pi(p\,|\,|\mathcal{C}|=n)$ obtained using the
proposed McMC for the scenario $\mathcal{S}2$ for the cluster size
values $p=10,\,35,\,50,\,70$ and corresponding trace plots.

\newpage 

\bibliography{PhysRev}

\end{document}